\documentclass[prl,twocolumn,superscriptaddress]{revtex4-2}
\usepackage{}
\usepackage{amssymb}
\usepackage{amsfonts}
\usepackage{bbm}
\usepackage{mathrsfs}
\usepackage{graphics,graphicx,epsfig,bm,amsmath,amsthm,amssymb}
\usepackage{bm}
\usepackage{bbm}
\usepackage{longtable}
\usepackage{multirow}
\usepackage{array}
\usepackage{color}
\usepackage[usenames,dvipsnames]{xcolor}
  
\usepackage{float}
\usepackage[a4paper,colorlinks=true,
linkcolor=blue,citecolor=blue,
pdfauthor={ },
pdftitle={ },
pdfsubject={ },
pdfkeywords={ }]{hyperref}

\begin{document}

\title{Experimental study of quantum coherence decomposition and trade-off relations in a tripartite system }

\author{Zhe Ding}
\email{These authors contributed equally to this work.}
\affiliation{Hefei National Laboratory for Physical Sciences at the Microscale and Department of Modern Physics,
	University of Science and Technology of China, Hefei, 230026, China}
\affiliation{CAS Key Laboratory of Microscale Magnetic Resonance, University of Science and Technology of China, Hefei, 230026, China}

\author{Ran Liu}
\email{These authors contributed equally to this work.}
\affiliation{Hefei National Laboratory for Physical Sciences at the Microscale and Department of Modern Physics,
	University of Science and Technology of China, Hefei, 230026, China}
\affiliation{CAS Key Laboratory of Microscale Magnetic Resonance, University of Science and Technology of China, Hefei, 230026, China}

\author{Chandrashekar Radhakrishnan}
\email{These authors contributed equally to this work.}
\affiliation{Laboratoire ESIEA Numérique et Société, ESIEA, 9 Rue Vesale, Paris 75005, France}
\affiliation{New York University, 1555 Century Avenue, Pudong, Shanghai 200122, China}
\affiliation{NYU-ECNU Institute of Physics at NYU Shanghai, 3663 Zhongshan Road North, Shanghai 200062, China}

\author{Wenchao Ma}
\affiliation{Department of Chemistry, Massachusetts Institute of Technology, Cambridge, Massachusetts 02139, USA}

\author{Xinhua Peng}
\affiliation{Hefei National Laboratory for Physical Sciences at the Microscale and Department of Modern Physics, University of Science and Technology of China, Hefei, 230026, China}
\affiliation{CAS Key Laboratory of Microscale Magnetic Resonance, University of Science and Technology of China, Hefei, 230026, China}
\affiliation{Synergetic Innovation Center of Quantum Information and Quantum Physics, University of Science and Technology of China, Hefei, 230026, China}

\author{Ya Wang}
\affiliation{CAS Key Laboratory of Microscale Magnetic Resonance, University of Science and Technology of China, Hefei, 230026, China}
\affiliation{Synergetic Innovation Center of Quantum Information and Quantum Physics, University of Science and Technology of China, Hefei, 230026, China}

\author{Tim Byrnes}
\email{tim.byrnes@nyu.edu}
\affiliation{New York University Shanghai, 1555 Century Ave, Pudong, Shanghai 200122, China}  
\affiliation{State Key Laboratory of Precision Spectroscopy, School of Physical and Material Sciences, East China Normal University, Shanghai 200062, China}
\affiliation{NYU-ECNU Institute of Physics at NYU Shanghai, 3663 Zhongshan Road North, Shanghai 200062, China}
\affiliation{National Institute of Informatics, 2-1-2 Hitotsubashi, Chiyoda-ku, Tokyo 101-8430, Japan}
\affiliation{Department of Physics, New York University, New York, NY 10003, USA}

\author{Fazhan Shi}
\email{fzshi@ustc.edu.cn}
\affiliation{Hefei National Laboratory for Physical Sciences at the Microscale and Department of Modern Physics, University of Science and Technology of China, Hefei, 230026, China}
\affiliation{CAS Key Laboratory of Microscale Magnetic Resonance, University of Science and Technology of China, Hefei, 230026, China}
\affiliation{Synergetic Innovation Center of Quantum Information and Quantum Physics, University of Science and Technology of China, Hefei, 230026, China}

\author{Jiangfeng Du}
\email{djf@ustc.edu.cn}
\affiliation{Hefei National Laboratory for Physical Sciences at the Microscale and Department of Modern Physics, University of Science and Technology of China, Hefei, 230026, China}
\affiliation{CAS Key Laboratory of Microscale Magnetic Resonance, University of Science and Technology of China, Hefei, 230026, China}
\affiliation{Synergetic Innovation Center of Quantum Information and Quantum Physics, University of Science and Technology of China, Hefei, 230026, China}

\begin{abstract}
Quantum coherence is the most fundamental of all quantum quantifiers, underlying other well-known quantities such as entanglement, quantum discord, and Bell correlations. It can be distributed in a multipartite system in various ways ---  for example, in a bipartite system it can exist within subsystems (local coherence) or collectively between the subsystems (global coherence), and exhibits a trade-off relation.   In quantum systems with more than two subsystems, there are more trade-off relations, due to the various decomposition ways of the coherence.  In this paper, we experimentally verify these coherence trade-off relations in adiabatically evolved quantum systems using 
a spin system by changing the state from a product state to a tripartite entangled state.  
We study the full set of coherence trade-off relations between the original state, the bipartite product state, the tripartite product 
state, and the decohered product state. We also experimentally verify the monogamy 
inequality and show that both the quantum systems are polygamous except for the initial product state.  
We find that despite the different types of states involved, the properties of the state in terms of coherence and monogamy 
are equivalent.  This illustrates the utility of using coherence as a characterization tool for quantum states.  
\end{abstract}
\maketitle

\noindent{\it Introduction.--} 
Quantum coherence has been the focus of investigation in numerous fields such as quantum optics 
where the fundamental nature of coherence has been investigated using phase-space distributions and higher order correlation functions 
\cite{glauber1963coherent,sudarshan1963equivalence,scully1999quantum}.  
It was quantified recently in a quantum-information theoretic way \cite{baumgratz2014quantifying} and the modern view is 
that it is the broadest quantum properties and is at the root of various quantum quantifiers such as discord, entanglement,
EPR steering, and Bell correlations \cite{ma2019operational,adesso2016measures}.  A set of axioms were formally 
introduced which need to be satisfied by a coherence quantifier \cite{baumgratz2014quantifying}.  This gave rise to the field of resource theories of quantum coherence
\cite{winter2016operational,chitambar2016critical,streltsov2017colloquium,streltsov2017structure}, along with an explosion of interest in measurement of coherence 
\cite{shao2015fidelity,radhakrishnan2016distribution,napoli2016robustness,girolami2014observable,radhakrishnan_basis-independent_2019} 
and its applications \cite{streltsov2016entanglement, karpat2014quantum,radhakrishnan2017quantum2,radhakrishnan2017quantum}. 

Quantum coherence has some unique features not present in other quantifiers such as entanglement and quantum discord (for a review see \cite{streltsov2017colloquium}).  One feature is that coherence is a basis dependent property and hence the amount of coherence depends upon the chosen measurement basis.  Additionally, coherence can localize in a unipartite system as quantum superposition, or be present as correlations between different qubits\cite{radhakrishnan2016distribution}.  For example, in a bipartite entangled state 
$ (|00 \rangle + |11 \rangle)/\sqrt{2}$, the coherence is delocalized and cannot be attributed to any particular qubit.  
On the other hand, in a separable state $ |++ \rangle = (|0\rangle + |1 \rangle)(|0\rangle + |1 \rangle)/2$,
the coherence is localized within the qubits.  In fact, a maximally entangled state has only global coherence and no 
local coherence; meanwhile product states are the opposite.
This example illustrates the presence of a trade-off between the local 
and the global coherence in a quantum system. 
This trade-off is the simplest case and highlights the different complementary distributions of coherence in two qubits. 
In a multipartite system there are more possible distributions of coherence and hence other types of trade-off relations. 

\begin{figure}[t]
		\includegraphics[width=\linewidth]{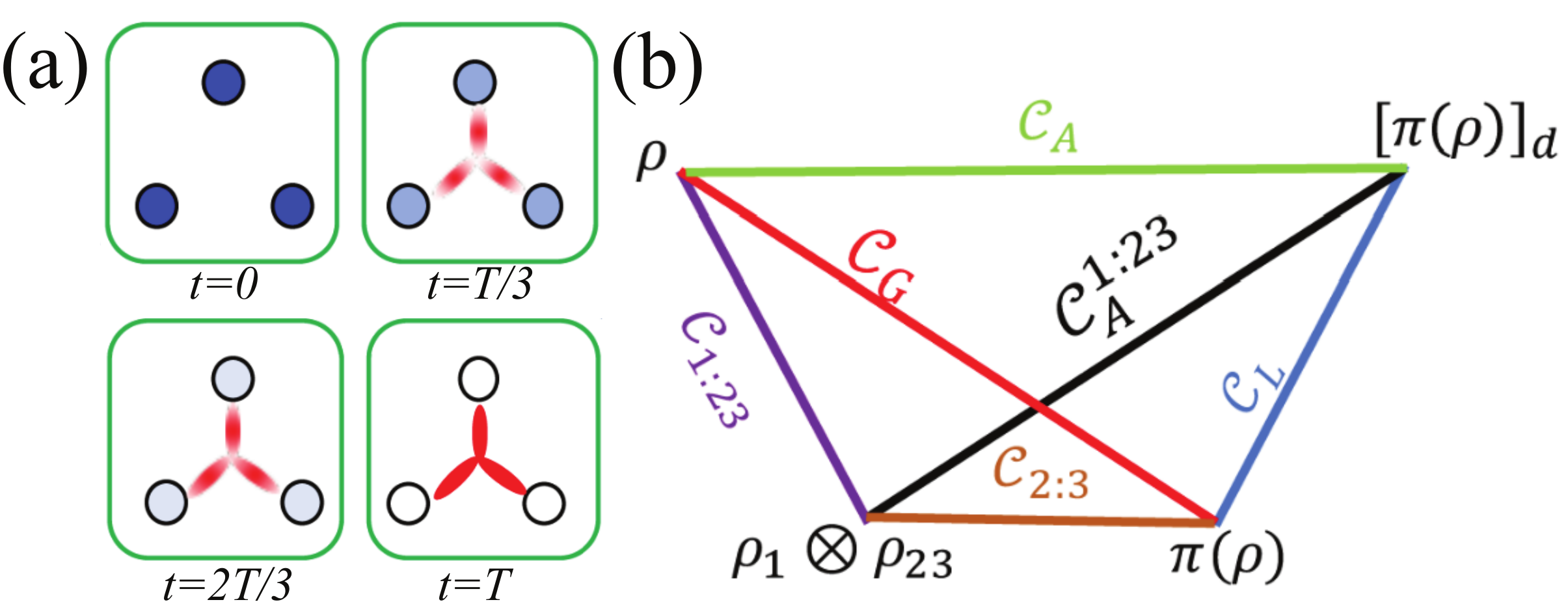} 
		\caption{(a) Quantum coherence trade-off is described in tripartite systems with circles representing the qubits. 
		The blue discs and red leaves represent the local coherence $C_L$ and the global coherence $C_G$. The strength of the color indicates the strength of the coherence. (b) A geometric 
		picture of different coherences. Coherences are shown as distances between two different density matrices. $\rho$ is the original density matrix; $\pi(\rho)$ and $[\pi(\rho)]_d$ are as described in (\ref{CgCl});  $\rho_{1} = {\rm tr}_{2,3} \: \rho$ and $\rho_{23} = {\rm tr}_{1} \: \rho$ are the reduced density metrices. }
		\label{fig1}
\end{figure}

In this Letter, we use a spin system to experimentally measure coherence and investigate the trade-off relations 
in tripartite systems.  We consider two different classes of quantum systems with two and three-body interactions. 
The quantum coherence is measured at different stages of an adiabatic evolution and various trade-off relations are verified.  
An example of this process is shown in Fig. \ref{fig1}(a), where initially the coherence is completely localized 
within the qubits.  Then, when the system is adiabatically evolved, it has both local and global coherence. At the end of the adiabatic evolution, the system has only global coherence.  In addition to verifying the trade-off relations and coherence distributions, we analyze the monogamy of coherence. Monogamy, first introduced in the context of entanglement \cite{coffman2000distributed,koashi2004monogamy}, implies that, when Alice and Bob are maximally entangled, they are impossible to be simultaneously entangled with a third party Charlie. This concept was later extended to quantum correlation \cite{giorgi2011monogamy,prabhu2012conditions} and quantum coherence  \cite{radhakrishnan2016distribution}.  We show that using the various coherence quantifiers and monogamy, one can reveal that despite the apparent differences between the two Hamiltonians, the decompositions of the coherence are in fact similar.  This illustrates the utility of our approach where coherence can be used to characterize a state to reveal hidden similarities between different systems.  

We note that several other works have examined quantum coherence experimentally recently \cite{exp1,exp2,wu2020quantum,yuan2020direct}.  
A coherence witness was introduced in  Ref. \cite{exp1} to detect the total coherence through a violation of Leggett-Garg type 
inequality \cite{Leggett_2008}.  Meanwhile the amount of coherence in a single photonic qubit was measured experimentally \cite{exp2} 
using the robustness of coherence \cite{napoli2016robustness}. These works studied coherence in unipartite systems and did not analyze the coherence decompositions, 
their distribution and the consequent trade-off relations in multipartite systems.

\noindent{\it Description of the quantum system.--}
We experimentally study tripartite quantum systems realized with nuclear spin qubits in this work.  Diethyl fluoromalonate molecules are used to perform the target adiabatic evolution via a Trotter decomposition. A complete description of the experimental set up and procedure is given in the Supplementary Materials.  In this work, we study two different tripartite quantum systems.  
The first system is an Ising model described by the Hamiltonian
\begin{equation}
H_{zz}(t) = \omega_z\sum_{i =1,2,3} S^z_i +  \omega_x \sum_{i =1,2,3} S^x_i +2J_2(t) 
                    \sum_{1 \leq i < j \leq 3} S^{z}_{i} S^{z}_{j},   
\label{Hamiltonian1}
\end{equation}
where $S^{z/x}_{i}$ is the nuclear spin in the $z/x$-direction and $J_{2}$ represents the two-body interaction strength with $\omega_{z} = -2$ 
being the magnetic field in the longitudinal direction.  A small transverse field $\omega_{x} = 0.1$ is provided to lift the degeneracy of the 
ground state so that the adiabatic evolution is possible.  Initially when we set $J_{2}=0$ and $\omega_{x} \ll \omega_{z}$, the ground state 
is nearly a separable state $|000 \rangle$.  The state is adiabatically evolved by increasing $J_{2}$ from 0 to $|\omega_{z}|$ in order to obtain a state close to $|W\rangle = (|001 \rangle + |010 \rangle + |100 \rangle)/\sqrt{3}$ at the 
end of the evolution \cite{peng_qpt2010}. The fidelity between the final ground state and $|W\rangle$ is 0.9978 while our experimental final state has a fidelity to $|W\rangle$ as high as 0.9578.

The second quantum system we consider has the following form:
\begin{equation}
H_{zzz}(t)  =  \omega_x \sum_{i =1,2,3} S^x_i + 4 J_{3}(t) S^{z}_{1} S^{z}_{2}S^{z}_{3},  
\label{Hamiltonian2}    
\end{equation}
where $J_{3}$ is the three-body interaction strength which varies from 0 to 5 during the adiabatic evolution.  The corresponding initial and final ground states are $|--- \rangle$ and, in the sense of zero-order perturbation, 
$|G\rangle = (|001\rangle + |010\rangle + |100\rangle  + |111\rangle)/2$, respectively. The fidelity between the final ground state and $|G\rangle$ is 0.9996 while our experimental final state has a fidelity to $|G\rangle$ as high as 0.9661. The final state has 
both bipartite and tripartite coherences. For both the quantum systems, the coherence is measured at each stage of the evolution using quantum tomography methods (see Supplementary Materials). 

 \noindent{\it Quantifying coherence.--}
 To measure coherence we use the square root of quantum version of the Jensen-Shannon divergence (QJSD) \cite{lin1991divergence,briet2009properties,majtey2005jensen,lamberti2008metric}
 \begin{equation}
 \mathcal{D} (\rho,\sigma) 
          = \sqrt{ \frac{1}{2} \left[\mathcal S_r(\rho \| (\rho+\sigma)/2) + \mathcal S_r(\sigma \|  (\rho+\sigma)/2) \right] }.
\label{QJSDmeasure}          
\end{equation}
Here $\rho$ and $\sigma$ are two density matrices of the same dimensionality and 
$\mathcal S_r(\rho_{1} \| \rho_{2}) = {\rm tr} \rho_{1} \log(\rho_{1}/\rho_{2})$ is the quantum relative entropy.  Using this measure, the total coherence in the system is
\begin{equation}
\mathcal{C}_{T} (\rho) \equiv  \mathcal{D}(\rho, \rho_{d}),
\label{totalcoherence}
\end{equation}
where $\rho$ is the density matrix and $\rho_{d} = \sum_{k} \langle k| \rho |k \rangle |k \rangle \langle k|$ is the diagonal density matrix 
with $|k \rangle$ representing the eigenstates of $S^{z}_{j}$. 
The global and local coherence are defined respectively as \cite{radhakrishnan2016distribution}
\begin{equation}
\mathcal{C}_G(\rho)  \equiv  \mathcal D(\rho,\pi(\rho));  \qquad
\mathcal{C}_{L}   \equiv    \mathcal{D}(\pi(\rho),[\pi(\rho)]_{d}).  
\label{CgCl}
\end{equation}
Here $\pi(\rho) \equiv \otimes_{i} \rho_{i}$, where $\rho_{i} = {\rm tr}_{\forall j \neq i} \: \rho$ and footnotes $i$, $j$ are indices of subsystems. Footnote $d$ indicates the diagonal part of the density matrix in the $S^z $-basis.    
In terms of the coherence trade-off, the more relevant quantity is the absolute coherence  defined as
\begin{equation}
\mathcal{C}_{A} (\rho) \equiv \mathcal{D}(\rho, [\pi(\rho)]_{d}), 
\label{absolutecoherence}
\end{equation}
which is different from the total coherence $\mathcal{C}_{T}$ and is the total amount of coherence in the product basis.  The reference state $ [\pi(\rho)]_{d} $  for absolute coherence contains neither coherence or correlations between the subsystems, while the reference state for total coherence $ \rho_{d} $ can potentially contain classical correlations.
Since our measure $\cal D(\rho,\sigma)$ satisfies the triangle inequality for a multipartite system up to five qubits according to numerical studies  \cite{lamberti_metric_2008}, we have the trade-off relation (see Fig. \ref{fig1}(b))
\begin{equation}
\mathcal{C}_{A} \leq \mathcal{C}_{L} + \mathcal{C}_{G}.
\label{tradeoffabsolute}
\end{equation}
The total coherence $\mathcal{C}_T$ does not satisfy this trade-off relation since the reference state $ \pi (\rho) $ is used.  

One of the interesting aspects of tripartite systems is that coherence can be distributed in different ways.  It is well-known that for
entanglement, GHZ and W states are two different classes of tripartite entangled states \cite{dur2000three}.  The entanglement 
in a GHZ state is genuinely tripartite, whereas in a W state, the entanglement is bipartite in nature.  In this context, it is interesting to examine 
the coherence additionally in a bipartite fashion. To this end, we evaluate the coherence between qubit 1 and the bipartite block 23
\begin{equation}
\mathcal{C}_{1:23} \equiv  \mathcal{D}(\rho,\rho_{1} \otimes \rho_{23}),  \quad  \mathcal{C}_{2:3} \equiv \mathcal{D}(\rho_{23}, \rho_{2} \otimes  \rho_{3}).
\label{asymmetriccoherencedistribution1}
\end{equation}
Here $\mathcal{C}_{1:23}$ measures the coherence between qubit $1$ and the bipartite block $23$ and $\mathcal{C}_{2:3}$ is the coherence in the 
bipartite block $23$.  We note that including $\rho_1$ does not make any difference to the global coherence, and 
$\mathcal{C}_{2:3} = \mathcal{D}( \rho_1 \otimes \rho_{23}, \pi(\rho))$; the proof is given in the Supplementary Materials. 
Complementary to the coherence $\mathcal{C}_{1:23}$ we have the contribution
\begin{equation}
\mathcal{C}_{A}^{1:23} \equiv \mathcal{D}(\rho_{1} \otimes \rho_{23}, [\pi(\rho)]_{d}).
\label{asymmetriccoherencedistribution2}
\end{equation}
Based on these coherence distributions we have the following trade-off relations
\begin{equation}
\mathcal{C}_{A}  \leq \mathcal{C}_{1:23} + \mathcal{C}_{A}^{1:23};  \qquad
\mathcal{C}_{A}^{1:23}  \leq \mathcal{C}_{2:3} + \mathcal{C}_{L}.
\label{asymmetrictradeoffrelations}
\end{equation}
Since  $\mathcal{C}_{1:23}$ and $\mathcal{C}_{2:3}$ are global coherences, they give another trade-off relation
\begin{equation}
\mathcal{C}_{G} \leq \mathcal{C}_{1:23} + \mathcal{C}_{2:3}. 
\label{tradeoffglobalcoherence}
\end{equation}
The four equations in  (\ref{tradeoffabsolute}), (\ref{asymmetrictradeoffrelations}) and (\ref{tradeoffglobalcoherence})
correspond to the four triangles that are present in Fig. \ref{fig1}(b).They are also four trade-off relations that can be verified for the generated tripartite states.  

\begin{figure}[t]
	\includegraphics[width=\linewidth]{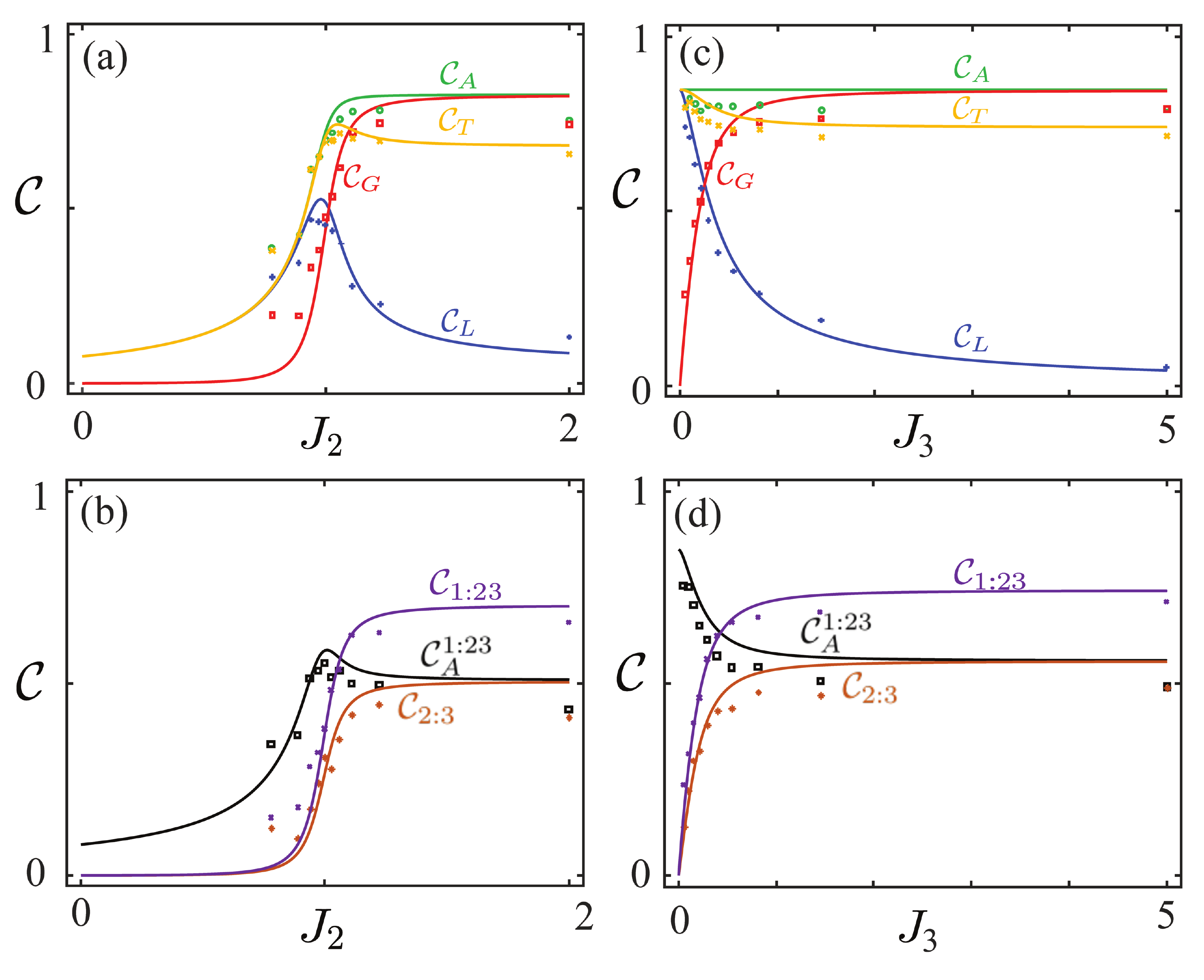} 
	\caption{The variation of different coherences as shown for the Hamiltonians (a),(b) $H_{zz}$ and (c),(d)  $H_{zzz}$, as a function of the interaction parameters.  The experimental data is shown by the points and the lines show the theoretically obtained results.}
	\label{fig2}
\end{figure}
\begin{figure}[h!]
	\includegraphics[width=\linewidth]{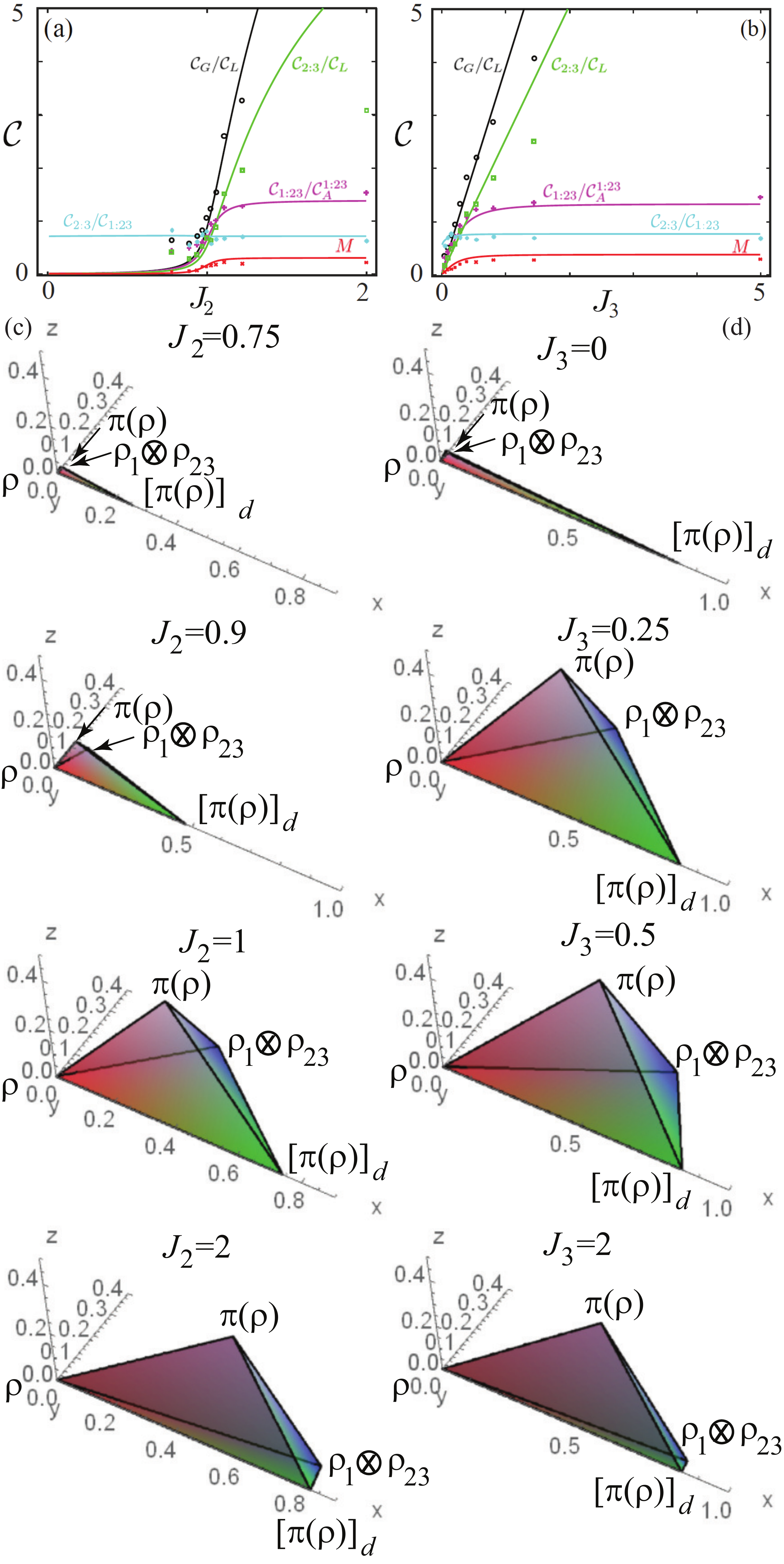} 
	\caption{The ratio between the different pairs of coherences and monogamy of coherence are shown in 
		(a)  Hamiltonian $H_{zz}$ and (b) Hamiltonian $H_{zzz}$ as a function of interaction parameters.  The points
		represent the experimental data and the solid lines correspond to the theoretical calculation. (c),(d) Geometric plots of the coherence in three dimensional Euclidean space for the $H_{zz}$ and $H_{zzz}$ respectively, for the values of interaction parameters as marked (top to bottom).  The lengths of the edges are taken to be the coherence contributions as shown in Fig. \ref{fig1}(b).  The coordinates of the state $ \rho $ is $ (0,0,0)$; $ [ \pi (\rho)]_d $ is $ ( {\cal C}_A, 0,0 ) $; $ \pi(\rho) $ is $ ( {\cal C}_G \cos \theta, {\cal C}_G \sin \theta, 0)$; $ \rho_1 \otimes \rho_{12} $ is $ ({\cal C}_{1:23} \cos \phi, {\cal C}_{1:23} \sin \phi \cos \xi, {\cal C}_{1:23} \sin \phi \sin \xi) $, where the angles are chosen to match the coherences as in Fig. \ref{fig1}(b). 	
	}
	\label{fig4}
\end{figure}
\noindent {\it Coherence trade-off.--} 
The variation of the coherence contributions during the evolution is shown in Fig. \ref{fig2}.  For Hamiltonian  $H_{zz}$, at $J_{2}=0$, the interactions are turned off and the ground state is a product state. It is locally rotated from the state $|000 \rangle$ ($\mathcal{C}_{L} = \mathcal{C}_{G} =0$), due to the transverse field $ \omega_x $ which  induces a small local coherence.  We observe that there are two regions, 
$J_{2} \in [0,1)$ with $\mathcal{C}_{L}$ and  $\mathcal{C}_{G}$ increasing and $J_{2} \in [1,2]$ with $\mathcal{C}_{L}$ 
decreasing and $\mathcal{C}_{G}$ increasing. The crossover at $J_{2}=1$ corresponds to a quantum phase transition in a spin system with two-body interactions \cite{peng_qpt2010}. For the Hamiltonian $H_{zzz}$, at $J_{3}=0$, the ground state is $|--- \rangle$, a coherent product state and hence $\mathcal{C}_{G}=0$ and $\mathcal{C}_{L}$ is 
maximal. At $J_{3}=5$, the ground state is nearly $|G \rangle$ for which $\mathcal{C}_{L}=0$ and $\mathcal{C}_{G}$ is maximal. 
The two distinct regions of  $H_{zzz}$ are $J_{3} \in [0,0.25)$ ($\mathcal{C}_{L} > \mathcal{C}_{G}$) and $J_{3} \in [0.25,5]$ 
($\mathcal{C}_{L} < \mathcal{C}_{G}$) with the crossover at $J_{3} =0.25$. They are related to a critical point at $J_3 = \omega_x$ for a spin system with three-body interaction in the thermal dynamic limit \cite{peng_qpt2010,igloi_conformal_1987,penson_conformal_1988,penson_phase_1982,igloi_critical_1983,baxter_exact_1973,pachos_three-spin_2004}. We note that there are regions where 
$\mathcal{C}_{G} > \mathcal{C}_{T}$ for both $H_{zz}$ and $H_{zzz}$.  This is due to our definition of global coherence, where all correlations between the qubits are broken by forming a product state, whereas in the definition of total coherence, classical correlations can be present in the decohered state. This verifies that  $\mathcal{C}_{A}$ is the more appropriate quantity in the context of trade-off relations.  

To visualize the expected trade-off relations according to 
(\ref{totalcoherence}), (\ref{tradeoffabsolute}) and (\ref{tradeoffglobalcoherence}), we look 
at the  ratios  $\mathcal{C}_{G}/\mathcal{C}_{L}$, $\mathcal{C}_{2:3}/\mathcal{C}_{L}$, $\mathcal{C}_{1:23}/\mathcal{C}^{1:23}_{A}$ 
and $\mathcal{C}_{2:3}/\mathcal{C}_{1:23}$ using both experimental data and the corresponding theoretical results as shown in Fig. \ref{fig4} .  
We observe three types of trade-off behavior corresponding to complete, partial and no trade-off.  For ratios of $\mathcal{C}_{G}$ and $ \mathcal{C}_{2:3} $ to $\mathcal{C}_{L}$, there is a complete trade-off between these quantities, since there is a complete exchange from locally to collectively distributed coherence.   Meanwhile the comparison of $\mathcal{C}_{2:3} $ and $\mathcal{C}^{1:23}_A $ to  $\mathcal{C}_{1:23} $  only results in a partial trade-off, where the ratios saturate to a finite value. In these cases, since both quantities in the ratio are types of global coherence, the ratios saturate to these particular values decided by large  $ J_2, J_3 $.  
 For Hamiltonian $ H_{zz} $, the ratio $\mathcal{C}_{2:3}/\mathcal{C}_{1:23}$ remains constant throughout.  We attribute this to the fact that for this Hamiltonian, there is a complete qubit symmetry, such that the entangled component of the state is always of the form of a W-state.  Hence when comparing two types of correlation-type coherences, although the amount of coherence both become small as $ J_2 \rightarrow 0 $, their ratio remains the same.  

The different kinds of coherences can also be visualized geometrically as shown in  Fig. \ref{fig4}(c),(d).  Here, we plot the various coherences by assigning them Euclidean distances in three dimensional space.   From the results we find that while $H_{zz}$ and $H_{zzz}$ have different kinds of interactions, their coherence distributions evolve similarly.  The general behavior is that the states $ \pi (\rho) $ and $ \rho_1 \otimes \rho_{12} $ start in the vicinity of $ \rho $, then eventually move to a location near $ [\pi (\rho)]_d $, along different trajectories. The primary difference between the two Hamiltonians is that $H_{zzz}$ always has a constant $ {\cal C}_A $, hence the size of the tetrahedron is of the same order, whereas for $H_{zz}$ the tetrahedron starts from a point.  However, apart from the overall magnitude of the coherence, the distribution of coherences are remarkably similar for both cases.  

\noindent {\it Monogamy of Coherence.--} 
The monogamy of coherence describes the trade-off between the bipartite and tripartite global coherences of a three-body quantum system \cite{radhakrishnan2016distribution}.  
We can quantify the monogamy according to $ M =  \mathcal{C}_{1:2} + \mathcal{C}_{1:3} - \mathcal{C}_{1:23} $, where $M > 0$ corresponds to a polygamous system and $M \leq 0$ to a monogamous system. 
The monogamy of coherence for the two Hamiltonians are shown in Fig. \ref{fig4}(a),(b). 
We find that the quantum systems are polygamous for every value of the interaction parameter except for the the initial value $ J_2, J_3 = 0 $.  This points to the fact that for both the quantum systems, the most dominant form of global coherence is the bipartite global coherence. 
 Since the coherences $\mathcal{C}_{1:23}$ and $\mathcal{C}_{2:3}$ are global 
 coherences, it is only natural that they are related to $C_{G}$, the total global coherence as explained in (\ref{tradeoffglobalcoherence}). This confirms the picture provided by Fig. \ref{fig4}(c),(d), that the coherence generated in the two Hamiltonians is of the same type.  This arises fundamentally because of the similar nature of the $ |W \rangle $ and $ |G \rangle $ state, which both have an bipartite-like entanglement structure.  

\noindent {\it Conclusions.--} 
We extended the notion of coherence trade-offs introduced in Ref. \cite{radhakrishnan2016distribution}
and experimentally studied all the trade-offs that are possible with the four point decompositions as shown in 
Fig. \ref{fig1}(b).  Each point in the diagram corresponds to removing a coherence contribution.  For example, the state 
$\pi(\rho)$ removes all the inter-qubit coherence and the state $[\pi(\rho)]_{d}$ removes all the coherence including that 
lying within the qubits.  Since we are dealing with a tripartite system, we further performed a bipartite decomposition 
where the coherence between the site $1$ and bipartite block $23$ is removed.  Our results point to the fact that the 
trade-off relations are generic behavior and are always obeyed as we move from a separable state to an entangled state. The trade-off behavior is also consistent with approaches where coherence is considered a resource, and coherence is converted into different forms \cite{wu2018experimental}, which may have different sensitivities to decoherence \cite{radhakrishnan2017quantum,PhysRevA.102.012403}. We also examined the distribution of global coherence using the property of monogamy of coherence and it was found that both the states were polygamous except when the interactions were turned off.   
The characterization of a quantum system through the coherence distribution diagrams in Fig. \ref{fig4}(c),(d) was shown to be an effective tool to visualize the quantum state.  It is interesting to note that while $H_{zz}$ and $H_{zzz}$
have different types of interactions and different initial states, the final quantum states have similar quantum properties.    
Hence, using the coherence decompositions and the trade-off relations, one can gain insight into the essential character of a given state, which may not be obvious simply by examining the wavefunction.  


\begin{acknowledgments}
The researchers at USTC are supported by the National Key Research and Development Program of China (Grants No. 2018YFA0306600 and 
2016YFA0502400), the National Natural Science Foundation of China (Grants No. 81788101, 91636217, 11722544, 11761131011, and 31971156),
the CAS (Grants No. GJJSTD20200001, QYZDY-SSW-SLH004 and YIPA 2015370), the Anhui Initiative in Quantum Information Technologies 
(Grant No. AHY050000), the National Youth Talent Support Program. TB and RC are supported by the Shanghai Research Challenge Fund; New York University Global Seed Grants for Collaborative Research; 
National Natural Science Foundation of China (61571301,D1210036A); the NSFC Research Fund for International Young Scientists 
(11650110425,11850410426); NYU-ECNU Institute of Physics at NYU Shanghai; the Science and Technology Commission of Shanghai 
Municipality (17ZR1443600); the China Science and Technology Exchange Center (NGA-16-001); and the NSFC-RFBR Collaborative 
grant (81811530112). 
\end{acknowledgments}

\bibliographystyle{apsrev}
\bibliography{ref}

\end{document}


\title{Experimental study of quantum coherence decomposition and trade-off relations in a tripartite system: supplementary material}

\author{Zhe Ding}
\email{These authors contributed equally to this work.}
\affiliation{Hefei National Laboratory for Physical Sciences at the Microscale and Department of Modern Physics,
	University of Science and Technology of China, Hefei, 230026, China}
\affiliation{CAS Key Laboratory of Microscale Magnetic Resonance, University of Science and Technology of China, Hefei, 230026, China}

\author{Ran Liu}
\email{These authors contributed equally to this work.}
\affiliation{Hefei National Laboratory for Physical Sciences at the Microscale and Department of Modern Physics,
	University of Science and Technology of China, Hefei, 230026, China}
\affiliation{CAS Key Laboratory of Microscale Magnetic Resonance, University of Science and Technology of China, Hefei, 230026, China}

\author{Chandrashekar Radhakrishnan}
\email{These authors contributed equally to this work.}
\affiliation{Laboratoire ESIEA Numérique et Société, ESIEA, 9 Rue Vesale, Paris 75005, France}
\affiliation{New York University, 1555 Century Avenue, Pudong, Shanghai 200122, China}
\affiliation{NYU-ECNU Institute of Physics at NYU Shanghai, 3663 Zhongshan Road North, Shanghai 200062, China}

\author{Wenchao Ma}
\affiliation{Department of Chemistry, Massachusetts Institute of Technology, Cambridge, Massachusetts 02139, USA}

\author{Xinhua Peng}
\affiliation{Hefei National Laboratory for Physical Sciences at the Microscale and Department of Modern Physics, University of Science and Technology of China, Hefei, 230026, China}
\affiliation{CAS Key Laboratory of Microscale Magnetic Resonance, University of Science and Technology of China, Hefei, 230026, China}
\affiliation{Synergetic Innovation Center of Quantum Information and Quantum Physics, University of Science and Technology of China, Hefei, 230026, China}

\author{Ya Wang}
\affiliation{CAS Key Laboratory of Microscale Magnetic Resonance, University of Science and Technology of China, Hefei, 230026, China}
\affiliation{Synergetic Innovation Center of Quantum Information and Quantum Physics, University of Science and Technology of China, Hefei, 230026, China}

\author{Tim Byrnes}
\email{tim.byrnes@nyu.edu}
\affiliation{New York University Shanghai, 1555 Century Ave, Pudong, Shanghai 200122, China}  
\affiliation{State Key Laboratory of Precision Spectroscopy, School of Physical and Material Sciences, East China Normal University, Shanghai 200062, China}
\affiliation{NYU-ECNU Institute of Physics at NYU Shanghai, 3663 Zhongshan Road North, Shanghai 200062, China}
\affiliation{National Institute of Informatics, 2-1-2 Hitotsubashi, Chiyoda-ku, Tokyo 101-8430, Japan}
\affiliation{Department of Physics, New York University, New York, NY 10003, USA}

\author{Fazhan Shi}
\email{fzshi@ustc.edu.cn}
\affiliation{Hefei National Laboratory for Physical Sciences at the Microscale and Department of Modern Physics, University of Science and Technology of China, Hefei, 230026, China}
\affiliation{CAS Key Laboratory of Microscale Magnetic Resonance, University of Science and Technology of China, Hefei, 230026, China}
\affiliation{Synergetic Innovation Center of Quantum Information and Quantum Physics, University of Science and Technology of China, Hefei, 230026, China}

\author{Jiangfeng Du}
\email{djf@ustc.edu.cn}
\affiliation{Hefei National Laboratory for Physical Sciences at the Microscale and Department of Modern Physics, University of Science and Technology of China, Hefei, 230026, China}
\affiliation{CAS Key Laboratory of Microscale Magnetic Resonance, University of Science and Technology of China, Hefei, 230026, China}
\affiliation{Synergetic Innovation Center of Quantum Information and Quantum Physics, University of Science and Technology of China, Hefei, 230026, China}

\maketitle

\section{Proof of $\mathcal C_{(1)2:3}=\mathcal C_{2:3}$}
One of the important distributions of quantum coherence is $C_{1:23}$ which is the coherence between the qubit $1$ and the bipartite block
$23$.  A complementary distribution is $C_{(1)2:3}$ which is the global coherence in the $\rho_{1} \otimes \rho_{23}$ system.  Below we prove 
that for the QJSD-based measure of quantum coherence $C_{(1)2:3} = C_{2:3}$.  

\begin{theorem}
$C_{(1)2:3}$ is equal to the global coherence $C_{2:3}$ in the system.  
\begin{proof}
The coherence $C_{(1)2:3}$ in terms of the QJSD based coherence measure is 
\begin{equation}
C_{(1)2:3} = \sqrt{ \mathcal{J}(\rho_{1} \otimes \rho_{23}, \rho_{1} \otimes \rho_{2} \otimes \rho_{3}) }.
\end{equation}
For two independent quantum systems $\rho$ and $\sigma$, the additivity of the von Neumann entropy leads to 
\begin{equation}
\mathcal S(\rho \otimes \sigma) = \mathcal S(\rho) + \mathcal S(\sigma).
\end{equation}
Due to this additivity, the QJSD obeys the restricted additivity \cite{majtey2005jensen} as given below:
\begin{equation}
\mathcal{J}(\rho \otimes \sigma_{1}, \rho \otimes \sigma_{2}) = \mathcal{J} (\sigma_{1}, \sigma_{2}),
\end{equation}
and so consequently we find that 
\begin{equation}
C_{(1)2:3} = \sqrt{ \mathcal{J} (\rho_{23}, \rho_{2} \otimes \rho_{3}) } = C_{2:3}.
\end{equation}
\end{proof}
\end{theorem}

\section{Analysis of ground states}
In this work, we investigate two different tripartite quantum systems by adiabatically evolving them.  In the first quantum 
system we have only two-body interactions and the Hamiltonian of the system reads:
\begin{equation}
H_{zz}(t) = \omega_z\sum_{i =1,2,3} S_i^z +  \omega_x \sum_{i =1,2,3} S_i^x +2J_2(t) 
\sum_{1 \leq i \leq j \leq 3} S_i^z S_j^z .
\label{H1}                    
\end{equation}
The Hamiltonian of the second quantum system with three-body interactions is 
\begin{equation}
H_{zzz}(t)  =  \omega_x \sum_{i =1,2,3} S_i^x + 4 J_{3}(t) S_1^z S_2^z S_3^z.
\label{H2}
\end{equation}
\begin{figure*}
	\centering
	\includegraphics[height=11cm]{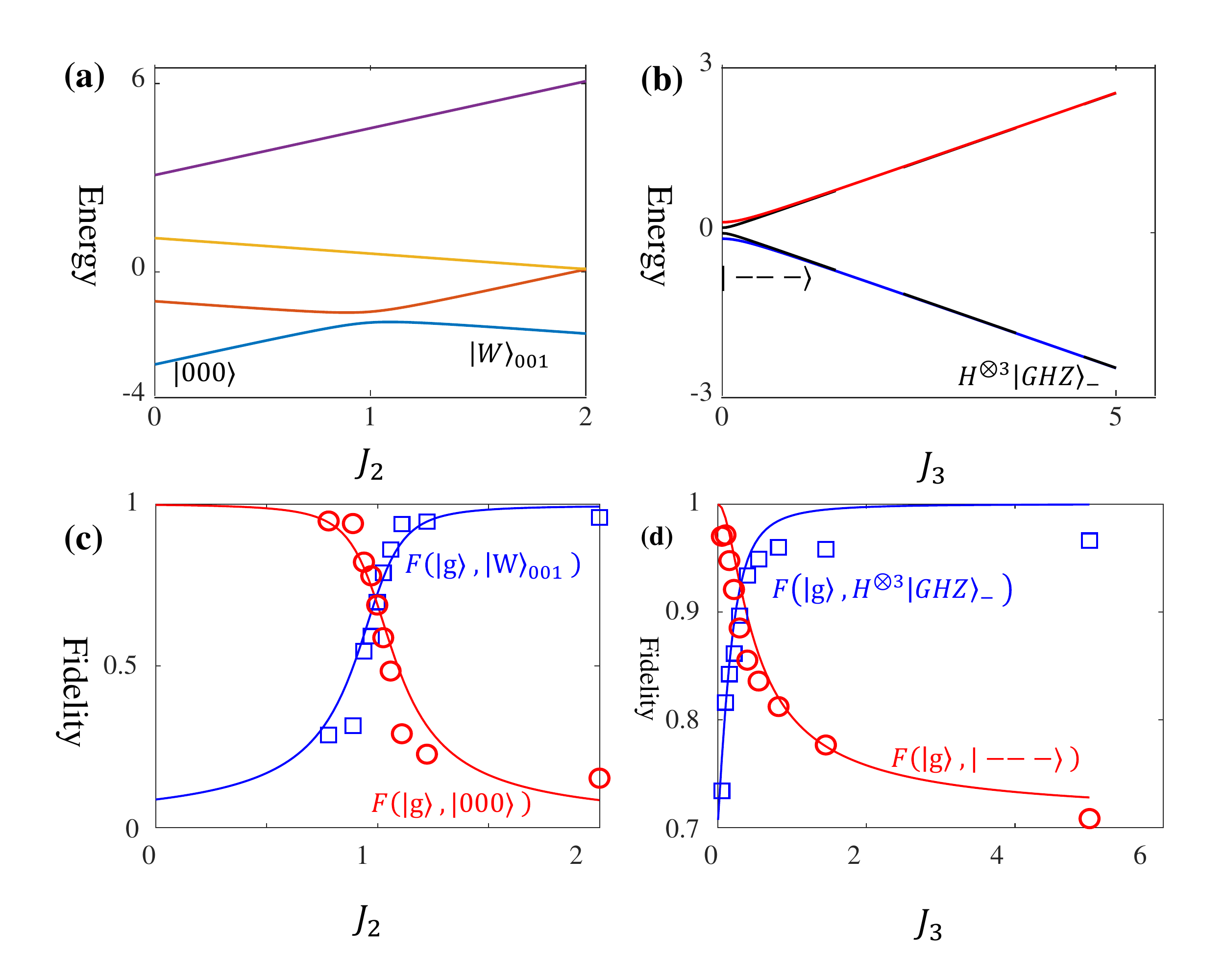}
	\caption{(a, b) The energy levels. (c, d) The fidelity between ground states and goal states. The solid lines show the numerical results. The circles and squares show the experiment results. }\label{figs1}
\end{figure*}
For the Hamiltonian $H_{zz}$, during the experiment, we assume $\omega_z=-2$, $\omega_x=0.1 \ll |\omega_z|$ and vary 
$J_2$ from $0$ to $2$.  The Hamiltonian is symmetric under permutation of spins.  When there is no perturbation, that is when 
$\omega_x=0, J_2=0$, the ground state is $|000 \rangle$.   We analyze the quantum state under the symmetric basis 
$\{|000\rangle, |W_{001}\rangle,|W_{110}\rangle, |111\rangle\}$ where
\begin{eqnarray}
|W_{001}\rangle&=&(|001\rangle+|010\rangle+|100\rangle)/\sqrt{3} \nonumber, \\
|W_{110}\rangle&=&(|110\rangle+|101\rangle+|011\rangle)/\sqrt{3} \nonumber.
\end{eqnarray}
The ground states are analyzed using perturbation theory. Defining $H_0=\omega_z\sum_{i =1,2,3} S_i^z +2J_2(t) \sum_{1 \leq i < j \leq 3} S_i^z S_j^z $ and $V=\omega_x \sum_{i =1,2,3} S_i^x $ 
where $V$ is the perturbation term.  The initial ground state under zero order perturbation is $|000\rangle$. At the end of the evolution, 
$\omega_z, J_2 \gg \omega_x$ under zero perturbation, the ground state is $|W\rangle_{001}$.  The corresponding energy level diagram is 
shown in Fig. \ref{figs1} (a). To compare the fidelity of the experimental state at $J_{2} =2$,  we can calculate the ground state up to the first order
perturbation\cite{MQM}
\begin{eqnarray}
|g\rangle & \approx & |g^{(0)}\rangle+\sum_{k\neq g}\frac{V_{kg}}{E_g^{(0)}-E_k^{(0)}} |k^{(0)}\rangle \nonumber \\
& = & |g^{(0)}\rangle + \frac{\omega_x}{\omega_z}|W\rangle_{110}-\frac{\sqrt 3}{2} \frac{\omega_x}{2J_2+\omega_z}|000\rangle
\end{eqnarray}
Here, $|g^{(0)}\rangle=|W\rangle_{001}$ is the zeroth order approximation of the ground state. 
The fidelity between $|W\rangle_{001}$ and final ground state in the sense of first order perturbation is : 
\begin{equation}
F(|g\rangle,|W\rangle_{001}) \approx 1/\left(1+\left(\frac{\omega_x}{\omega_z}\right)^2+\left(\frac{\sqrt 3}{2} \frac{\omega_x}{2J_2+\omega_z}\right)^2\right)
\end{equation}

We also numerically diagonalize the Hamiltonian directly and calculate the fidelities. In Fig. \ref{figs1}(c), we show the numerically calculated fidelities $F(|g\rangle,|W\rangle_{001})$  through a blue solid line and the experimental data fidelities 
through blue squares.  From the figure, we can see that the numerically calculated final fidelity is $0.9978$ and the experimental fidelity is 
as high as $0.9578$.

Next, we analyze the ground state of Hamiltonian $H_{zzz}$.  The initial Hamiltonian 
$J_{3}=0$ is $H_{zzz}(0)=\omega_x \sum_{i =1,2,3} S_i^x$ and the initial ground state is  $|---\rangle$, where 
$|-\rangle =  (|0 \rangle - |1 \rangle)/\sqrt{2}$.  At the end of the adiabatic evolution the final Hamiltonian is obtained 
when $J_{3}=5$.  For the final Hamiltonian $J_{3} \gg \omega_{x}$, the unperturbed part of the Hamiltonian is 
$H_0= 4 J_{3}(t) S_1^z S_2^z S_3^z$ and the perturbation term is $V=\omega_x \sum_{i =1,2,3} S_i^x$.
The states $|W\rangle_{001}$ and $|111\rangle$ expands a degenerated ground state subspace in the non-perturbed Hamiltonian.  
Since the transition term between these two states in $H_{zzz}$ is zero, we need to use first-order perturbation theory, in which 
the projection of the approximated ground state to the degenerate subspace is the solution to the secular equation:
\begin{equation}
\sum_{\nu \in g} c_{g\nu} \left( \sum_{n \notin g}\frac{V_{g\mu,n}V_{n,g\nu}}{E_g^{(0)}-E_n^{(0)}}-\delta_{\mu\nu}\Delta E_g \right)=0
\label{secularequation}
\end{equation} 
In the above equation, $\Delta E_g$ is the perturbation of the ground state's energy, $g$ represents the degenerate ground state subspace, 
$\mu,\nu$ are the labels of the basis in the subspace. We label $|W\rangle_{001}$ and $|111\rangle$ as $g1$ and $g2$ respectively. 
By solving Eq. (\ref{secularequation}), we find:
\begin{eqnarray}
|g\rangle & \approx & \frac{\sqrt 3}{2} |W\rangle_{001} +\frac{1}{2} |111\rangle \nonumber \\
& = & H^{\otimes 3}|GHZ\rangle_- \\ 
& = & |G\rangle \nonumber
\end{eqnarray}
The energy levels of this evolution are plotted in Fig. \ref{figs1}(b).  To estimate the final state, we calculate the first order 
perturbation of the ground state and include the contribution from the other two energy levels.  
\begin{eqnarray}
|g\rangle & \approx & |G\rangle+\sum_{n\notin g}\sum_\mu\frac{V_{n,g\mu}c_{g\mu}}{E_g^{(0)}-E_n^{(0)}}|n^{(0)}\rangle \nonumber \\
& = & |G\rangle -\frac{\omega_x}{J_3}\left(\frac{3}{4}|000\rangle+\frac{3\sqrt 3}{4}|W\rangle_{110}\right)
\end{eqnarray}
Using this we can estimate the final state's fidelity, 
\begin{equation}
F(|g\rangle,|G\rangle)\approx 1/\left(1+\left(\frac{3\omega_x}{2J_3}\right)^2\right)
\end{equation}
We also numerically diagonalize the Hamiltonian directly and calculate the fidelities. The numerically calculated value of the fidelity is shown through blue lines in Fig. \ref{figs1} (d), 
and the fidelity of the final state is $0.9996$. The fidelities of experimental data are displayed as blue squares among which the final fidelity is $0.9661$. 

\begin{figure*}
	\centering
	\includegraphics[height=14cm]{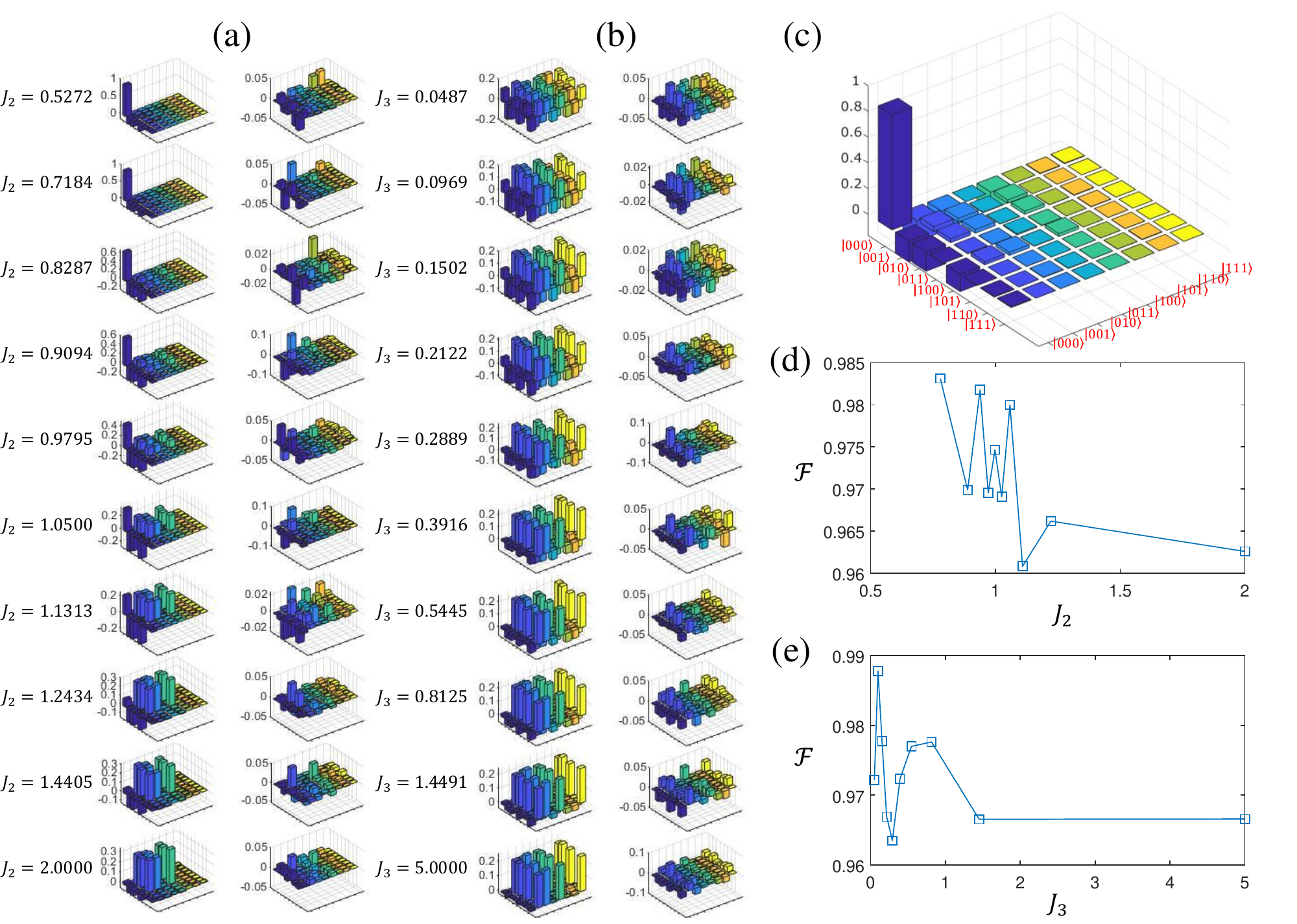}
		
	\caption{Tomography results of the quantum states. (a,b) show the tomography results of all the experimental states for $H_{zz}$ and $H_{zzz}$, the first columns show the real part of the density matrices while the second columns the imaginary part. (c) shows the basis of the density matrices displayed in (a,b). (d,e) show the fidelities between experimental states and ground states. }\label{tomo}
\end{figure*}

The tomography of the experimental states is shown through Fig. (\ref{tomo}) (a) and (b)
for the Hamiltonians $H_{zz}$ and $H_{zzz}$ respectively and in Fig. (\ref{tomo}) (c) the corresponding basis is shown for reference.  
In Fig. \ref{tomo} (d) and (e) we display the fidelities between the experimental states and the corresponding ground states. 

\section{Experimental protocol}
\begin{figure}[tb]
	\begin{center}
		\includegraphics[width=1\linewidth]{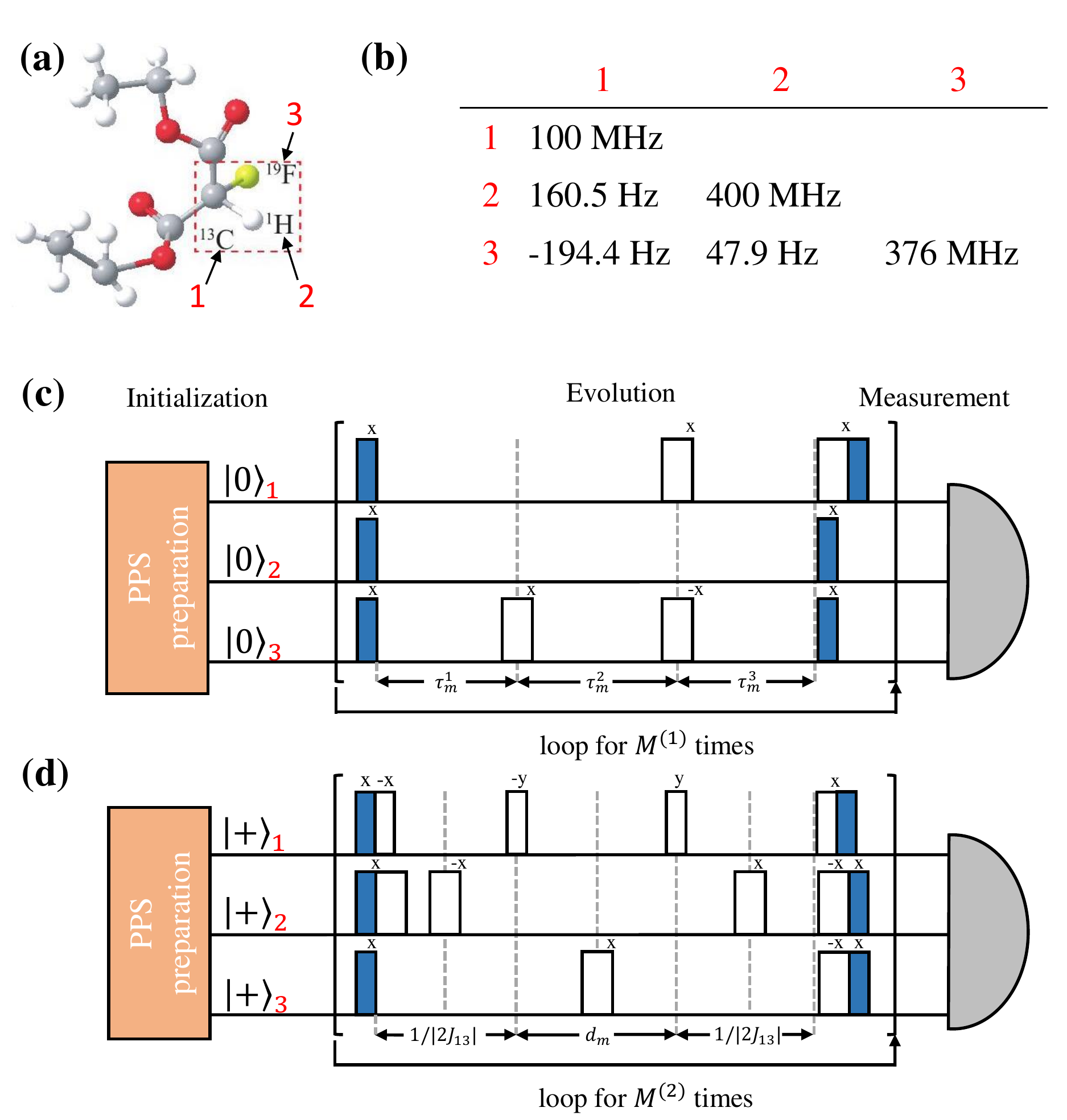} 
		\caption{ a) Molecular structure of Diethyl fluoromalonate. The three nuclear spins $^{13}$C, $^{1}$H and $^{19}$F adopted in the experiment are labeled. The corresponding qubit index is also marked in red for each nuclear spin with a black arrow.  b) Parameters of the natural Hamiltonian of the three-spin system are shown in this table.  The diagonal terms are the values of the chemical shift and the off-diagonal terms represent the scalar coupling between the different nuclei. c) and d) shows the schematic diagram explaining the experimental procedure for $H_{zz}$ and $H_{zzz}$ respectively.  In the initialization part, we prepare the system from a pseudopure state (PPS) into the designed Hamiltonian's ground state. In the evolution part, a discrete refocusing scheme is used to perform two Hamiltonians. Combining with Trotter expansion, we can perform adiabatic evolution under any interaction. The wide and narrow unfilled pulses represent $\pi$ and $\pi/2$ pulses respectively while the rotation axes are labeled above each pulse. The filled pulses represent a rotation of $\omega_x \tau/2$, where $\tau$ means the length of each Trotter slice. The measurement part is carried out using quantum state tomography. }
		\label{scheme}
	\end{center}
\end{figure}

In the experiment, we use  diethyl fluoromalonate molecules dissolved in ${}^{2}$H-labeled chloroform as the three-qubit spin system.
The molecular structure of diethyl fluoromalonate is shown in Fig. \ref{scheme} (a). The three nuclear spins  ${}^{13}$C, ${}^{1}$H and ${}^{19}$F 
in the molecule acts as the qubits.  The natural Hamiltonian of the system is 
\begin{equation}
H_{\rm{spin}} =  \sum_{i=1,2,3} 2 \pi \delta_{i} S_{i}^{z}  + \sum_{1 \leq i \leq j \leq 3} 2 \pi J_{ij} S_{i}^{z} S_{j}^{z}
\end{equation}
where $S_{i}^{z}$ is the nuclear spin in the $z$-direction, $\delta_{i}$ is the chemical shift of the nuclear spin and $J_{ij}$ is the 
coupling between the $i$-th and the $j$-th nucleus as given in Fig. \ref{scheme} (b).  The NMR experiment was carried out on a Bruker Avance III
400 MHz (9.4 T) spectrometer at 303K.   In the first step of the experiment, a pseudopure state (PPS) of the form 
$\rho = (1 - \mu) I/8 + \mu |\psi \rangle \langle \psi |$ is prepared from thermal equilibrium state using a line-selective method\cite{peng2001preparation}, where 
$|\psi \rangle$ is an arbitrary pure state.  Here the mixing parameter $\mu \approx 10^{-5}$ and  $I$ denotes 
the $8 \times 8$ identity matrix.  The adiabatic pathway is numerically optimized to generate the desired ground state.  The schematic diagram 
of the sequence to fulfill $H_{zz}$ and $H_{zzz}$ are shown in Fig. \ref{scheme} (c) and (d). At each stage, the corresponding density matrices 
are reconstructed using tomographic techniques.

\section{Experimental details of the refocusing scheme}\label{scheme_detail}
Experimentally, the adiabatic evolution is performed in discrete steps.  The evolution of each segment $U^{(k)}_{exp}(t_m)$, is a Trotter expansion 
of the ideal one $U^{(k)}_{ide}(t_m)$, which can be expressed as
\begin{eqnarray}
	U^{(k)}_{ide}(t_m)&=&e^{-i[H^{(k)}_x+H^{(k)}_z(t_m)]\tau^{(k)}} \nonumber \\ 
	&=&e^{-i H^{(k)}_x \tau^{(k)}/2} e^{-i H^{(k)}_z(t_m)\tau^{(k)}}e^{-i H^{(k)}_x \tau^{(k)}/2}+O(\tau^3)\nonumber \\
	&=&U^{(k)}_{exp}(t_m)+O(\tau^3)
	\label{trotter}
\end{eqnarray}
where $k\in\{1,2\}$ labels two Hamiltonians, i.e. $H^{(1)}=H_{zz}, H^{(2)}=H_{zzz}$. $\tau^{(k)}$ is the interval of each step and $m\in [0,M]$ is the index of each step. 
We use a refocusing scheme to achieve each step in our work.  In this method, tuned pulses are applied during each Trotter slice, 
and the Hamiltonian in each short time period is accurately controlled.  

Two different quantum systems are experimentally studied in our work.  In the first system, we construct a tripartite Hamiltonian with 
identical two-body interactions as shown in Eq. (\ref{H1}).  The quantum system is adiabatically evolved by tuning the two qubit interaction 
strength adiabatically over the range $[0,2]$.   Experimentally, the adiabatic state transfer (ASP) is performed in discrete steps, 
such that $J_{2}(t)$ assumes discrete value $J_{2} (t_{m})$ with $m=0,...,M^{(1)}$.  At each time step, the evolution is generated 
using multipulse sequence $U^{(1)}_{exp}(t_m)$ using Trotter expansion formula as described in Eq. (\ref{trotter}).  The resulting 
Hamiltonian is 
\begin{eqnarray}
H^{(1)}_{x}  &=& \omega_x \sum_{i =1,2,3} S_i^x,\\
 H^{(2)}_{z}(J_2(t_m))   &=& \omega_z\sum_{i =1,2,3} S_i^z+2J_2(t_m) \sum_{1 \leq i \leq j \leq 3} S_i^z S_j^z \nonumber
\end{eqnarray}
A schematic description of the refocusing scheme is shown in Fig. \ref{scheme} (c) where the narrow unfilled rectangles 
denote $\pi/2$ pulses, and the wide ones show $\pi$ pulses.  By defining $d_{ij}=1/(2J_{ij})$, the width of filled pulse in (c) are all 
$\omega_x\tau^{(1)}/2$ and the radio-frequency offsets for three channels are set as 
$FQ1_m=\omega_z/(4J_2(t_m)d_{12}),\ FQ2_m=\omega_z/(4J_2(t_m)(d_{12}+d_{13}+d_{23}))\text{ and}\ FQ3_m=\omega_z/(4J_2(t_m)d_{23})$, 
the delays are $\tau^{(1)}_m=\frac{J_2(t_m)\tau^{(1)}}{\pi}\times(d_{12}+d_{23}),\ \tau^{(2)}_m=\frac{J_2(t_m)\tau^{(1)}}{\pi}\times(d_{12}+d_{13})$, 
and $\tau^{(3)}_m=\frac{J_2(t_m)\tau^{(1)}}{\pi}\times(d_{13}+d_{23})$. 

Next we consider the Hamiltonian of a tripartite quantum system with $J_{3}$ being the three-body interaction strength as shown in Eq. (\ref{H2}).
The interaction parameter $J_{3}$ is tuned adiabatically in the range $[0,5]$.  Again, we use a discrete refocusing scheme in which 
$J_{3}(t)$ is discretized into $t_{m}$, $m=0,...,M^{(2)}$.  The schematic diagram is shown in Fig. \ref{scheme} (d) in which the width of the 
filled pulse are all $\omega_{x} \tau^{(2)}/2$ and the delay $d_{m}  = \frac{ J_{3} (t_{m}) \tau^{(2)} }{\pi} \times d_{12}$.

From above, one can see that the unit of studied quantities like $J_2, \omega_z, \tau^{(i)}$ always cancel out when they 
come into the parameters of the experiment. This means that the units of them do not matter in the experiment, 
only the relative relations between them matter. So, they are in arbitrary units and we don't mention the unit in the main text. We use 0.7 and 0.4 as the value of $\tau^{(1),(2)}$ when we design the experimental sequences, to see the reason why these two values are chosen, please refer to the next section.

\section{Optimization of experimental parameters}

\begin{figure*}
	\begin{minipage}{1\linewidth}
		\centering
		\includegraphics[height=5.5cm]{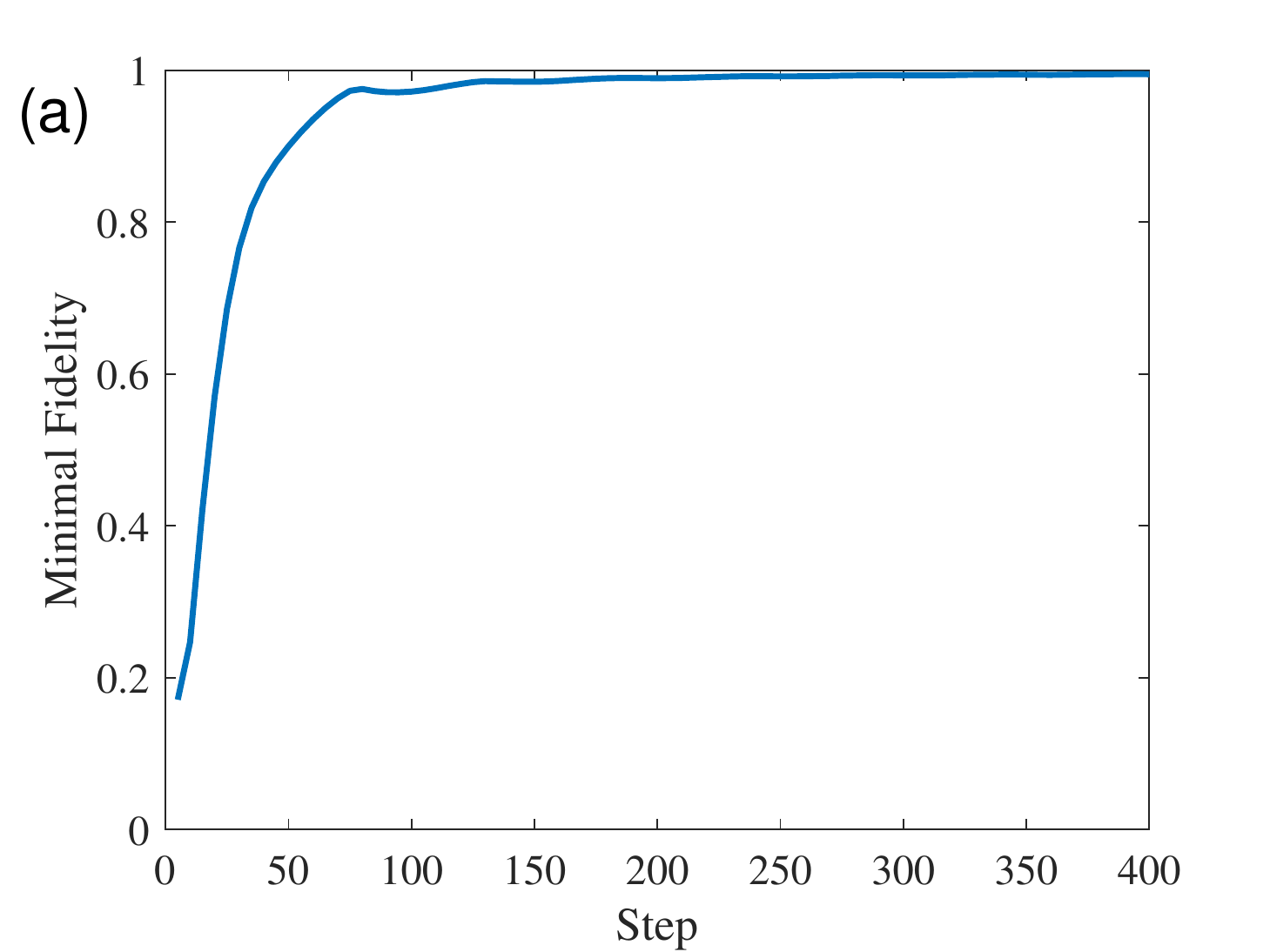}
		\centering
		\includegraphics[height=5.5cm]{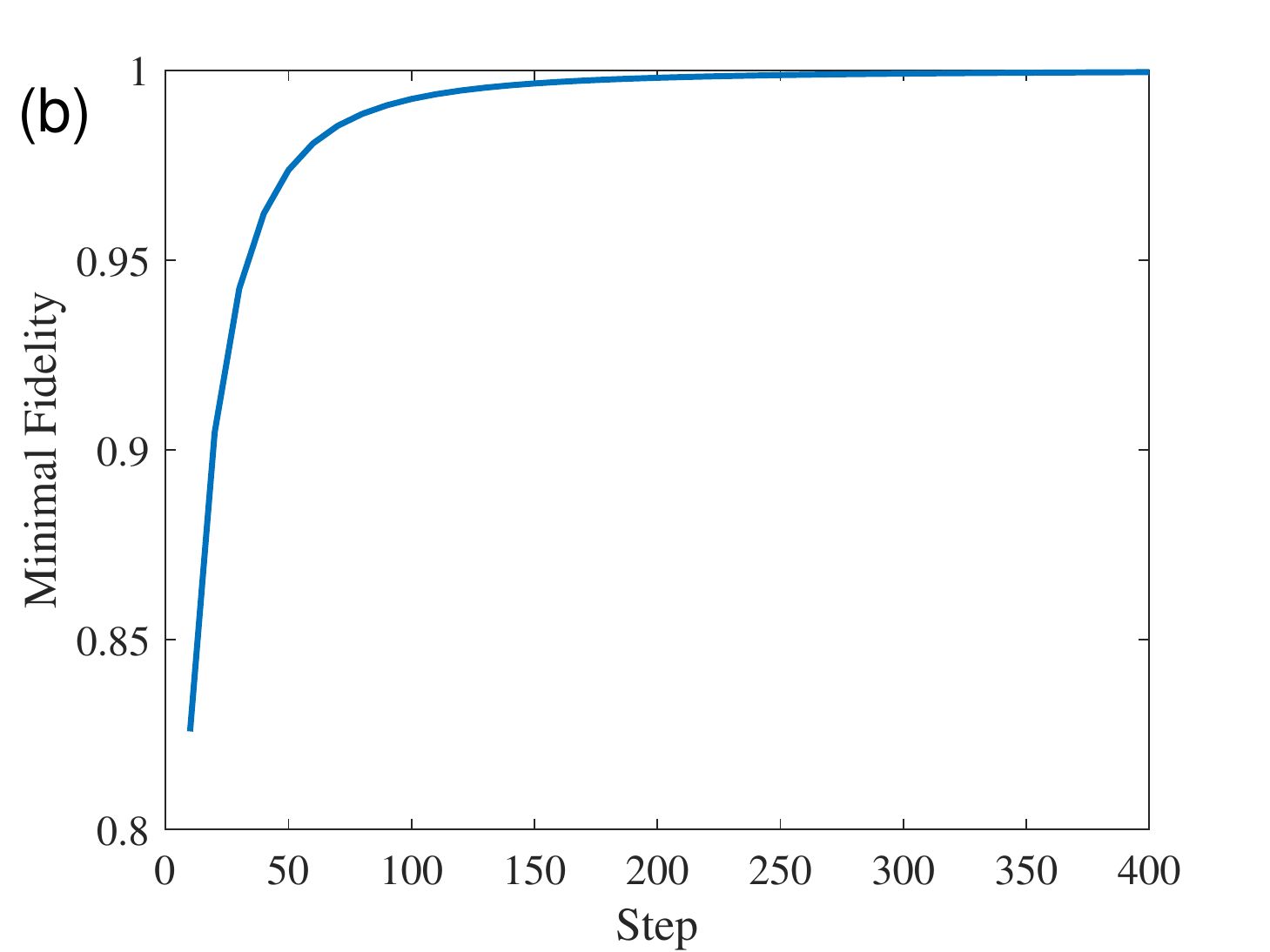}\\
		\centering
		\includegraphics[height=5.5cm]{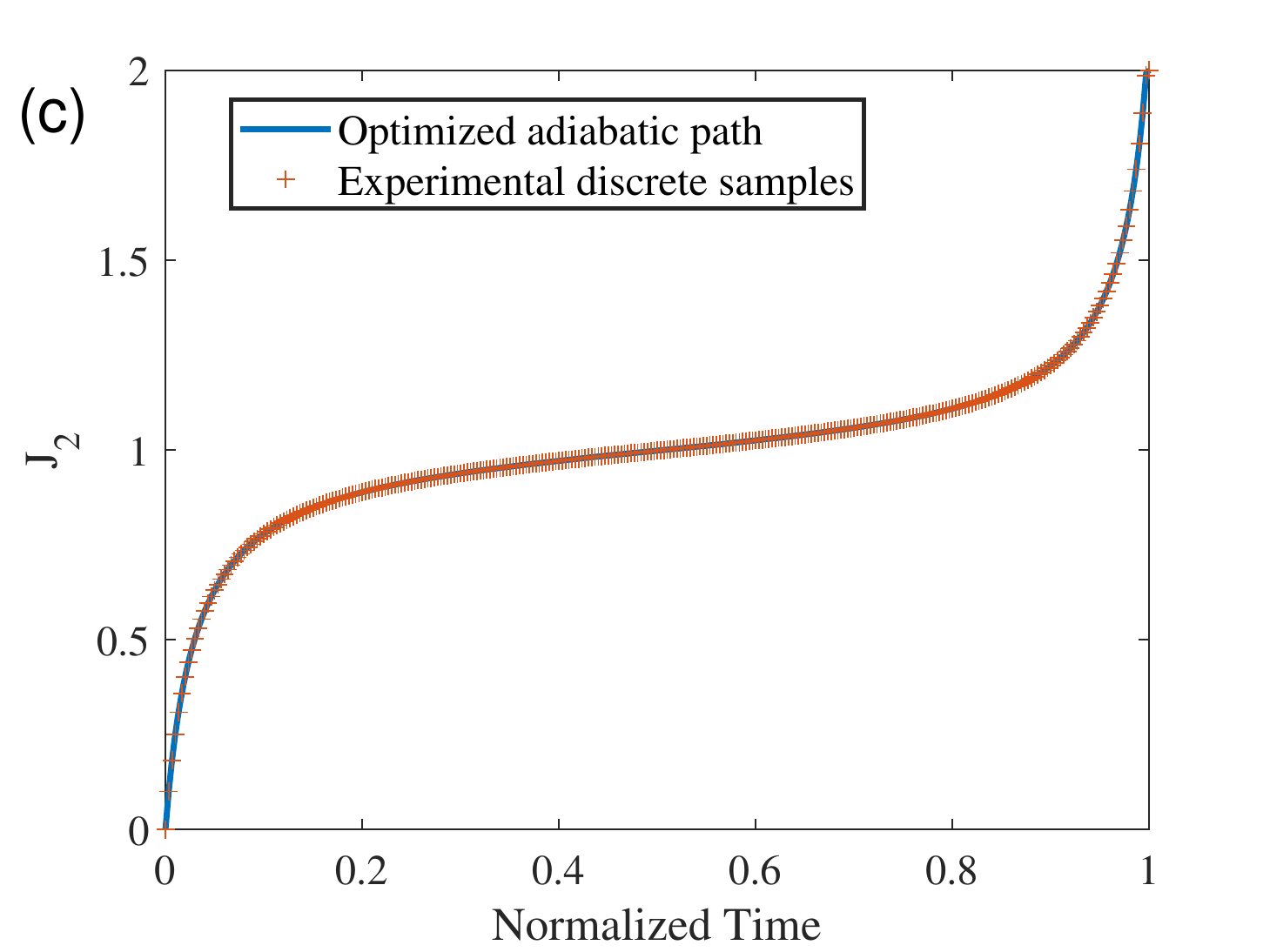}
		\centering
		\includegraphics[height=5.5cm]{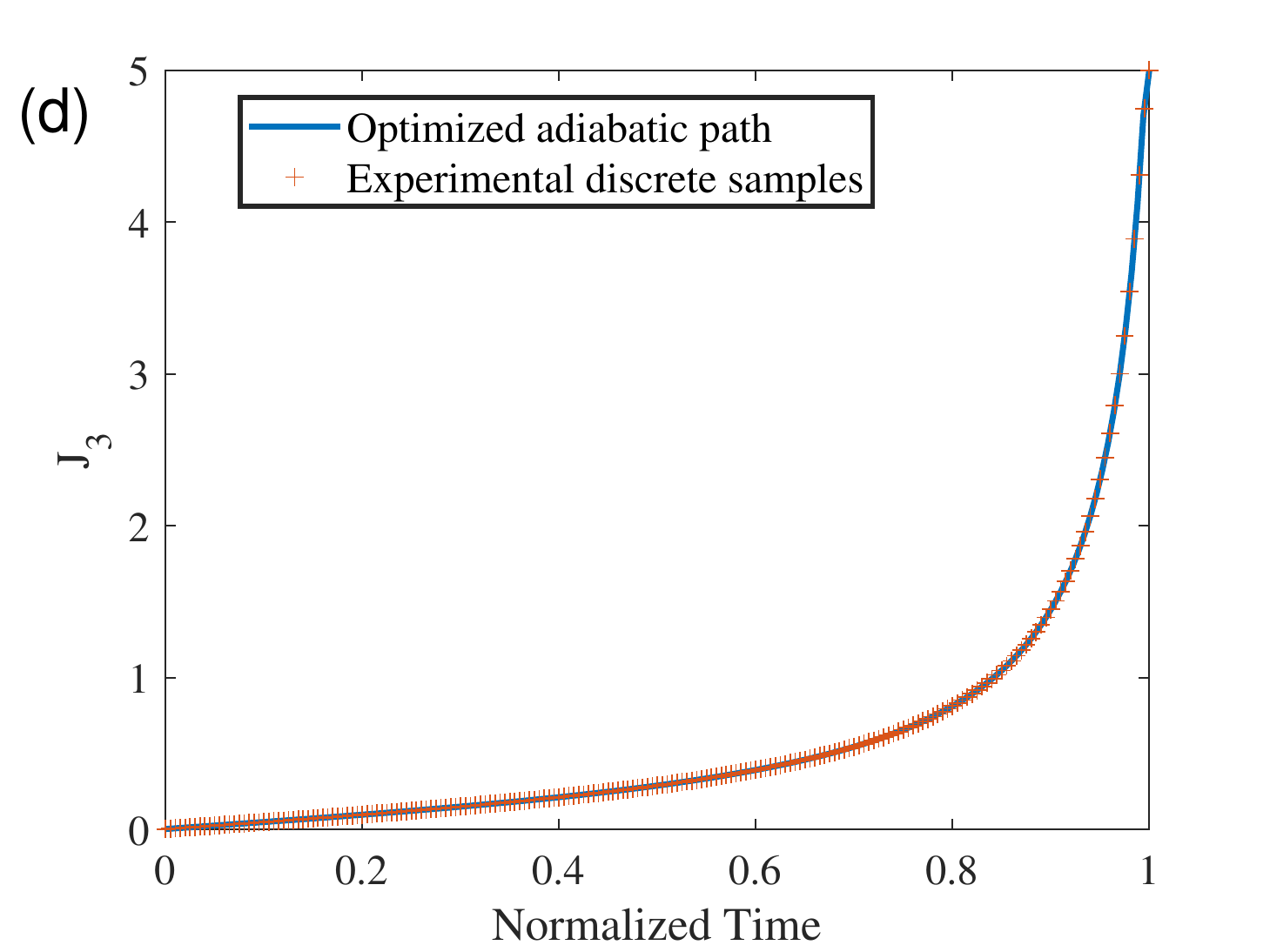}
	\end{minipage}
	\caption{(a, b) The minimum fidelity of generated states during the adiabatic process of $H^{(1),(2)}$ versus the number of steps. (c, d) The optimized adiabatic paths of $H^{(1),(2)}$ are shown as the blue solid lines, while the discrete points are sampled according to $M^{(1),(2)}$ for experimental implementing.}\label{subFig1}
\end{figure*}

In the previous section, we discussed about the adiabatic evolution.  As a result, $\tau$ should be kept small enough to maintain the 
fidelity $U^{(k)}_{ide}(t_m)$ and $U^{(k)}_{exp}(t_m)$ high for all $m\in[0,M]$, while it should not be too small that the number of discrete steps is large. 
We set $\tau^{(1),(2)}$ as $0.7,0.4$, respectively, as it guarantees that $f[U^{(k)}_{exp}(t_m),U^{(k)}_{ide}(t_m)]>99.9\%$.  
Here, $f(U_1,U_2)$ is the fidelity between $U_1$ and $U_2$, which is defined as $f(U_1,U2)=\frac{|Tr(U_1\times U_2')|^2}{d^2}$, and $d$ is 
the dimension of $U_{1,2}$. We then increase the number of steps and obtain the minimum fidelity of generated states during the adiabatic process by 
numerical simulation.  From the adiabatic paths obtained from numerical optimization, the linear interpolation method is employed and the corresponding 
discrete path can then be obtained by simulation.  According to the results shown in Fig.\ref{subFig1} (a,b), we finally set  $M^{(1)}=300$ and $M^{(2)}=200$, respectively, 
to make sure the generated states are kept as close as possible to the theoretical ground states.  The stepwise values of $J_2, J_3$ for implementing 
in the experiment are shown as the discrete points in Fig.\ref{subFig1}(c,d). Obviously, the discrete samples are denser when the gap between the 
two lowest energy levels decreases, and this is why the numerically optimized path can adjust speed of adiabatic transfer according to the structure of 
energy level and speedup the adiabatic process.

\bibliographystyle{apsrev}
\bibliography{refsupp}